\pdfoutput=1
\documentclass[12pt,aps,nofootinbib]{revtex4-2}

\usepackage[colorlinks=true,citecolor=blue,urlcolor=blue,linkcolor=blue,pdfstartview=FitH,bookmarksopen]{hyperref}
\usepackage{amsmath,amssymb}
\usepackage{bm}
\usepackage{comment}
\usepackage[usenames,dvipsnames]{xcolor}
\usepackage{color,graphicx}
\renewcommand\d{\partial}
\newcommand\vareps{\varepsilon}

\newcommand\x{\mathbf{x}}
\newcommand\X{\mathbf{X}}
\newcommand\y{\mathbf{y}}
\renewcommand\u{\mathbf{u}}
\newcommand\q{\mathbf{q}}

\newcommand\dl{\overset{\leftarrow}\partial}
\newcommand\dr{\overset{\rightarrow}\partial}
\newcommand\Dlr{\overset{\leftrightarrow}D}

\newcommand\A{{\mathcal A}}
\newcommand\F{{\mathcal F}}

\newcommand\bA{A^b}

\newcommand\blambda{\lambda^b}

\newcommand\sfX{\mathsf{X}}
\newcommand\sfA{\mathsf{A}}

\newcommand\cA{A^c}
\newcommand\clambda{\lambda^c}

\newcommand\V{{\mathcal V}}
\newcommand\+\dagger

\renewcommand\>{\rangle}

\begin{document}

\title{Noncommutative gauge symmetry in the\\ fractional quantum Hall effect}
\author{Yi-Hsien Du, Umang Mehta, and Dam Thanh Son}
\affiliation{Kadanoff Center for Theoretical Physics, University of Chicago, Chicago, Illinois 60637, USA}

\begin{abstract}

We show that a system of particles on the lowest Landau level can be
coupled to a probe U(1) gauge field $\A_\mu$ in such a way that the
theory is invariant under a noncommutative U(1) gauge symmetry.  While
the temporal component $\A_0$ of the probe field is coupled to the
projected density operator, the spatial components $\A_i$ are best
interpreted as quantum displacements, which distort the interaction
potential between the particles.  We develop a Seiberg-Witten-type map
from the noncommutative U(1) gauge symmetry to a simpler version,
which we call ``baby noncommutative'' gauge symmetry, where the Moyal
brackets are replaced by the Poisson brackets. The latter symmetry
group is isomorphic to the group of volume preserving diffeomorphisms.
By using this map, we resolve the apparent contradiction between the
noncommutative gauge symmetry, on the one hand, and the particle-hole
symmetry of the half-filled Landau level and the presence of the mixed
Chern-Simons terms in the effective Lagrangian of the fractional
quantum Hall states, on the other hand.  We outline the general
procedure which can be used to write down effective field theories
which respect the noncommutative U(1) symmetry.

\end{abstract}

\maketitle

\section{Introduction}

The problem of the fractional quantum Hall effect is often formulated
in the limit of a single Landau level (e.g., the lowest Landau level
(LLL)), in which the nontriviality of the problem becomes most stark.
In this limit, the drift motion of a single electron in an external
potential can be captured by a Hamiltonian formalism in which the two
Cartesian coordinates $x$ and $y$ of the electrons do not commute with
each other.  It has been suggested that the quantum field theory
describing the fractional quantum Hall fluids has to be a
noncommutative (NC) field
theory~\cite{Susskind:2001fb,Polychronakos:2001mi,Hellerman:2001rj,Fradkin:2002qw,Cappelli:2009pn}.
However, noncommutativity of space in the first-quantized description
does not translate directly to a noncommutativity of space in the
second-quantized formalism.  Some progress in deriving noncommutative
field theories for the quantum Hall effect has recently been made for
bosonic quantum Hall states near filling factor
$\nu=1$~\cite{Dong:2020bkt}, based on previous works by Pasquier and
Haldane~\cite{Pasquier:1997aik} and Read~\cite{Read:1998dn}.

In this paper we try to address the question of whether the quantum
Hall states are governed by a noncommutative field theory.  Instead of
following a constructive route, we will base our approach on symmetry
constraints.  We show that the LLL electrons can be coupled to an
external probe which can be identified with a NC U(1) gauge potential
$\A_\mu$ in a way that partition function of the theory is invariant
under NC U(1) gauge transformation.  However, $\A_\mu$ is not the
usual electromagnetic probe.  While the scalar potential $\A_0$ is
coupled simply to the projected density operator, the vector
components $\A_i$ do not couple to the charge current.  In fact,
$\A_i$ is best interpreted as a displacement, disturbing the shape of
the potential between two electrons.  So in addition to the NC gauge
symmetry, there is an additional symmetry where $\A_i$ is shifted by a
functions of time which are independent of space
[Eq.~(\ref{A-translation}) below].

Armed with the symmetry information, one can then proceed to a
construction of effective field theories describing various fractional
quantum Hall states.  The simplest way to guarantee the invariance of
the theory with respect to the NC U(1) symmetry is to try to promote
field theories of the fractional quantum Hall effect to noncommutative
field theories.
However, one immediately encounters serious problems along the way.
First, one sees that the naive noncommutative version of the Dirac
composite fermion theory now breaks particle-hole symmetry.
Furthermore, the mixed Chern-Simons term, crucial for the construction
of the effective field theory for many quantum Hall states (the
``hierarchy states'')~\cite{Wen:1992uk} , does not have an obvious
noncommutative extension.

We solve this problem by developing a map, inspired by the
Seiberg-Witten (SW) map~\cite{Seiberg:1999vs}, which we term the ``SW'
map,'' which maps the noncommutative probe field $A_\mu$ into what we
call ``baby noncommutative'' (bNC) gauge field $\bA_\mu$.  In the
transformation laws of the bNC gauge field, the Moyal brackets are
replaced by the Poisson brackets. The bNC U(1) gauge group is
isomorphic to the group of volume-preserving diffeomorphisms (VPDs).
Thus the task of writing down an action invariant under the gauge
symmetry reduces to ensuring volume-preserving diffeomorphism
invariance, which can be accomplished by using an appropriate
geometric formalism, for example, the Newton-Cartan
formalism~\cite{Son:2013rqa,Du:2021pbc}.  In this way one can write
down a particle-hole-symmetric effective field theory for the
half-filled Landau level as well as a Chern-Simons theory with a
general $K$-matrix.

The paper is organized as follows.  In Sec.~\ref{sec:noncomm-review}
we review the basic formulas of noncommutative space and
noncommutative gauge theory.  In Sec.~\ref{sec:symmetry-LLL} we couple
the electrons on a single Landau level to a NC U(1) gauge field.  In
Sec.~\ref{sec:noncomm-EFT} we outline the ways one can construct
effective field theories which respect the NC U(1) gauge symmetry.
Section~\ref{sec:conclusion} contains final remarks.

\section{Noncommutative space and noncommutative gauge symmetry}
\label{sec:noncomm-review}

For convenience and further reference, in this Section we collect some
formulas related to noncommutative space and noncommutative gauge
symmetry.

Consider particles in a magnetic field $B=\d_1 A_2-\d_2 A_1$.
The gauge-invariant momenta are
\begin{equation}
  \hat p_i = -i (\d_i - iA_i) = \hat k_i - A_i (\hat x).
\end{equation}
They satisfy the commutation relation
\begin{equation}
  [\hat p_i, \, \hat p_j] = i \vareps_{ij} B .
\end{equation}
Projecting to the LLL effectively sets $p_i\approx 0$ (the kinetic
energy is $\hat p^2/2m$, LLL projection corresponds to taking $m\to
0$). This introduces the Dirac brackets between $x_i$:
\begin{equation}
  [\hat x^i, \, \hat x^j]_{\text{D}} =
  - [\hat x^i, \, \hat p_k] ([\hat p,\, \hat p])^{-1}_{kl}
  [\hat p_l,\, \hat x^j]
    = -i \ell^2 \vareps^{ij} .
\end{equation}
So effectively the particle lives in a noncommutative space, with the
noncommutative parameter $\theta^{ij}=\theta\vareps^{ij}$ with
\begin{equation}
  \theta = - \ell^2.
\end{equation}

Any operator in the Heisenberg algebra (i.e., can be expanded in
Taylor series over $\hat x_i$) can be put into correspondence it
\emph{Weyl symbol} by the following rule
\begin{equation}
  e^{iq_i \hat x^i} \rightarrow e^{iq_i x^i} .
\end{equation}
Taking all possible linear combinations of the exponential one can put
any function $f(x)$ into correspondence with an operator
$\hat f(\hat x)$ in a unique way.
(We will use the same letter for the operator and
its Weyl symbol, putting a hat on top of the the symbol for the
operator.)

The Moyal product between two functions $f(x)$ and $g(x)$, $f\star g$,
is defined so that the Weyl symbol of $\hat f \hat g$ is $f\star g$.
\begin{equation}
  f \star g(x) = \exp\left( \frac i2 \theta
  \vareps^{ij} \frac\d{\d x^i} \frac\d{\d y^j} \right)
  f(x) g(y) \biggl|_{y\to x}
  = f(x) \exp\left( \frac i2\theta \vareps^{ij} \dl_i \dr_j\right) g(x).
\end{equation}
The Moyal bracket is defined as
\begin{equation}
  \{\!\!\{ f, \, g \}\!\!\} = \frac 1i (f\star g - g\star f)
  = 2 f\sin \left( \frac 12\theta \vareps^{ij} \dl_i \dr_j\right)g,
\end{equation}
and is (up to the factor $-i$) the Weyl symbol of the commutator
$[\hat f,\, \hat g]$.  The factor of $-i$ makes sure that if $f$ and
$g$ are real functions then $\{f,\, g\}$ is also real.  To leading
order in $\theta$ the Moyal bracket becomes the Poisson bracket
\begin{equation}
  \{\!\!\{ f, \, g\}\!\!\} =  \theta \{f, \, g\} + O(\theta^2),
\end{equation}
where
\begin{equation}
  \{ f, \, g\} = \vareps^{ij}\d_i f \d_j g .
\end{equation}

We now consider noncommutative gauge symmetry.  For the purposes of this paper,
we can limit ourselves to the U(1) case and only to adjoint fields.
An adjoint field $\psi$ transforms under a noncommutative
$U(1)$ gauge transformation as
\begin{equation}
  \delta_\lambda \psi
  = \{\!\!\{ \psi,\, \lambda \}\!\!\} ,
\end{equation}
with $\lambda$ being the parameter of the infinitesimal gauge transform.
The covariant derivative is defined as
\begin{equation}
  D_\mu \psi =
  \d_\mu \psi + \{\!\!\{ \A_\mu ,\,  \psi \}\!\!\} ,
\end{equation}
and the gauge potential transforms as
\begin{equation}\label{deltaAmu-NC}
  \delta_\lambda \A_\mu = \d_\mu \lambda + \{\!\!\{ \A_\mu,\, \lambda \}\!\!\}. 
\end{equation}
The gauge invariant gauge field is
\begin{equation}
  \mathcal F_{\mu\nu} = \d_\mu \A_\nu - \d_\nu \A_\mu + \{\!\!\{ \A_\mu, \, \A_\nu\}\!\!\} .
\end{equation}

\section{Symmetry of the Landau level problem}
\label{sec:symmetry-LLL}

\subsection{Introduction of external probe fields}

The starting point of our discussion is the Hamiltonian describing a
system of $N$ particles on a single Landau level, interacting with
each other through a two-body potential $V(\x-\y)$,
\begin{equation}\label{H-projected}
  H = \sum_{\<ab\>} \hat {\mathcal V}( \hat\x_a - \hat \x_b)
  = \sum_{\<ab\>}
  \int\! \frac{d\q}{(2\pi)^2}\, \mathcal V(\q) e^{i\q\cdot(\hat \x_a -\hat \x_b) }, 
\end{equation}
where $\mathcal V(\q)$ is the projected version of the inter-particle
potential $V(\q)$
\begin{equation}
  \mathcal V(\q) = V(\q)  e^{-q^2\ell^2/2} L_n^2\left(\frac{q^2\ell^2}2\right),
\end{equation}
with $L_n$ being the $n$th Laguerre polynomial and $n$ is the quantum
number of the Landau level on which the electrons live (for a quick
derivation of Eq.~(\ref{H-projected}) see, e.g.,
Ref.~\cite{Yang:2013}).

We will couple the theory to a set of probe fields.  These probes are
functions in spacetime and act on all particles in the same ways.  The
first probe is the scalar potential, or the temporal component of the
electromagnetic field $A_0(\x)$.  The Hamiltonian now contains a one-body
term
\begin{equation}
  H = \sum_{\<ab\>} \hat{\mathcal V}( \hat\x_a - \hat \x_b)
  - \sum_a \hat \A_0(\hat \x_a),
\end{equation}
where, on the lowest Landau level, $\A_0(\q) = e^{-q^2\ell^2/4} A_0(\q)$.

Now we introduce two more probe fields which we call $u^x(\x)$ and
$u^y(\x)$, forming a vector $\u(\x)$.  These probes replace the
coordinates of the particles in the interaction potential $\V(\x_a, \x_b)$ by new coordinates
\begin{equation}
  \x \to \X = \x + \u(\x).
\end{equation}
The perturbed Hamiltonian is now
\begin{multline}
  \hat H = \sum_{\<ab\>} \hat \V (\hat \X_a - \hat \X_b)
  - \sum_a \hat \A_0(\hat \x_a)
  = \sum_{\<ab\>} \hat \V \left( \hat \x_a  + \hat \u(\hat \x_a)
  - \hat \x_b - \hat \u(\hat \x_b) \right)
   - \sum_a \hat \A_0(\hat \x_a) .
\end{multline}
Note that we do not shift the coordinates in the one-body term $\hat \A_0$.
The probes can also be made time-dependent: $\A_0=\A_0(t,\x)$,
$\u=\u(t,\x)$.

\subsection{Detour: time dependent unitary transformation}

Consider a system described by a Hamiltonian $\hat H$ (which, in
general can be time-dependent, $\hat H=\hat H(t)$). Suppose we know
how to find all solutions $|\psi(t)\>$ to the time-dependent
Schr\"odinger equation
\begin{equation}
  i \frac\d{\d t} |\psi\> = \hat H(t)| \psi\> .
\end{equation}
If now $\hat U$ is a (time-independent) unitary operator, and $\hat
H_U= \hat U \hat H\hat U^{-1}$, then $|\psi\>_U = U|\psi\>$ is the
solution to the Schr\"odinger equation with $\hat H$ replaced by $\hat
H_U$.

But one can also perform a time-dependent unitary transformation with
time-dependent $\hat U$.  It is easy to see that if one sets
\begin{equation}
  \hat H_U (t) = \hat U \hat H \hat U^{-1} + i\d_t \hat U \hat U^{-1}, 
\end{equation}
then the state $|\psi\>_U = U(t)|\psi\>$ still solves the
time-dependent Schr\"odinger equation.  Therefore the new Hamiltonian
is completely equivalent to the old Hamiltonian.  In particular, if
$\hat U(t)$ is not equal to 1 only in a finite time interval
$t_i<t<t_f$, then the evolution operator $T \exp \left[
-i \int_{t_i}^{t_f} dt\, H(t)\right]$ does not change when $H(t)$ is
replaced by $H_U(t)$.

We now consider an infinitesimal unitary transformation with $\hat U =
e^{i\hat\lambda}$ where $\hat\lambda\ll 1$.  The transformed
Hamiltonian is then
\begin{equation}
  \hat H_\lambda = \hat H + i[\hat \lambda, \, \hat H] - \d_t \hat\lambda.
\end{equation}

\subsection{Unitary transformation in the Landau-level problem}

We now act on our Hamiltonian describing $N$ particles noncommutative plane
with the following infinitesimal unitary transformation
\begin{equation}
  \hat U = \exp\left[ i\sum_a \hat\lambda(t, \hat \x_a) \right],
\end{equation}
where $\lambda(t,\x)$ is an infinitesimal function of $t$ and $\x$.
Note that all particles are subjected to the same unitary
transformation.

The transformed Hamiltonian is
\begin{equation}
  \hat H^\lambda = \hat U \hat H \hat U^{-1} + i\d_t \hat U \hat U^{-1}
  =  \sum_{\<ab\>} \V(\hat \X_a^\lambda- \hat \X_b^\lambda)
  - \sum_a \hat \A_0^\lambda(t, \hat\x_a),
\end{equation}
where
\begin{align}
  \hat \X^\lambda &
  = \hat \X + i [\hat \lambda, \, \hat \X],\\
  \hat \A_0^\lambda &= \hat \A_0 + i [\hat \lambda, \hat A_0] + \d_t \hat\lambda .
\end{align}

So the new Hamiltonian has the same form as the old Hamiltonian but
with the new values of the probe fields.  The transformation laws of
the corresponding Weyl symbols are
\begin{align}
  \X^\lambda & = \X + \{\!\!\{\X , \, \lambda \}\!\!\} , \label{X-transform} \\
  \A_0^\lambda &= \A_0 + \{\!\!\{ \A_0, \, \lambda \}\!\!\} + \d_t \lambda . 
\end{align}

Knowing how $X^i$ transform we can find how $u^i$ transforms.  We have
$\X = \x+\u$, so Eq.~(\ref{X-transform}) reads
\begin{equation}
  x^i + (u^i)^\lambda  =
  x^i +  u^i + \{\!\!\{ x^i + u^i, \, \lambda \}\!\!\} .
\end{equation}
But
\begin{equation}
  \{\!\!\{  x^i, \, \lambda \}\!\!\} = - \ell^2 \vareps^{ij}\d_j \lambda ,
\end{equation}
which means
\begin{equation}
  (u^i)^\lambda = u^i + \{\!\!\{ u^i, \, \lambda \}\!\!\}
  - \ell^2 \vareps^{ij} \d_j \lambda .
\end{equation}

Now if we introduce $\A_i$ so that
\begin{equation}
  u^i = -\ell^2 \vareps^{ij} \A_j ,
\end{equation}
then $\A_i$ transforms as
\begin{equation}
  \A_i^\lambda = \A_i + \{\!\!\{ \A_i, \, \lambda\}\!\!\} + \d_i \lambda .
\end{equation}
That means we can combine $\A_0$ with $\A_i$ into $\A_\mu$, all components
of which transform in the same way
\begin{equation}
  \delta_\lambda  \A_\mu = \{\!\!\{  \A_\mu,\, \lambda \}\!\!\} + \d_\mu \lambda  .
\end{equation}
This is exactly the gauge transformation of the noncommutative $U(1)$
gauge symmetry.

So the system of particles on a Landau level can be coupled to a set
of gauge potentials $\A_\mu$ so that the physics is invariant under
the noncommutative U(1) gauge transformation.  It follows that any
low-energy effective theory should also couple to the $\A_\mu$ probe
in a way that respects this invariance.  Note that while $\A_0$ is
related to the temporal component of the electromagnetic field, $\A_i$
is not at all the usual vector potential of electromagnetism.

Beside the noncommutative U(1) gauge symmetry, translational
invariance of the two-body potential term leads to a symmetry
\begin{equation}\label{A-translation}
  \A_i (t,\x) \to \A_i(t,\x) + \alpha_i(t),
\end{equation}  
where $\alpha_i(t)$ are functions of time only.  This additional
symmetry should also be respected by any low-energy effective field
theory.




\section{Noncommutative symmetry in effective field theory}
\label{sec:noncomm-EFT}

\subsection{Problem with particle-hole symmetry}

The effective field theory describing the half-filled Landau
level~\cite{Halperin:1992mh,Son:2015xqa} is a U(1) gauge theory which
involves, as dynamical degree of freedom, a composite fermion $\psi$
and an emergent gauge field $a_\mu$.  The effective action has a
$U(1)_a$ emergent gauge symmetry, and also the conventional
$U(1)_\text{em}$ gauge symmetry, encoded in the coupling of the theory
to external background gauge field $A_\mu$.

One now asks how one can couple the theory to the noncommutative probe
field $\A_\mu$.  One may expect that the field theory should be
promoted to a noncommutative field theory.  In this scenario the
$U(1)_a$ gauge symmetry would become a $U(1)_a$ noncommutative gauge
symmetry.  In order to ensure the time-dependent shift
symmetry~(\ref{A-translation}), one may require all fields to
transform in the adjoint representation of the U(1) NC gauge symmetry.


Such noncommutative field theory of the half-filled Landau level can
be constructed (see, e.g., Ref.~\cite{Gocanin:2021xto})
but one sees an
immediate problem.  Namely, the NC $U(1)_a$ gauge symmetry is in
conflict with the particle-hole (PH) symmetry (or $\mathcal{CT}$
symmetry) of the physics on a single Landau level.  To show that, let
us write down the transformation laws for the gauge fields under gauge
transformations~\cite{Son:2015xqa},
\begin{align}
	& U(1)_a^\text{NC} : \qquad \delta_\alpha a_\mu = \{\!\!\{  a_\mu,\, \alpha \}\!\!\} + \d_\mu \alpha , \label{a-NCtrans} \\
	& U(1)_\text{em}^\text{NC} : \qquad \delta_\lambda \A_\mu = \{\!\!\{  \A_\mu,\, \lambda \}\!\!\} + \d_\mu \lambda .
\end{align}
Under particle-hole symmetry the fields transform as follows:
\begin{align}
	A_0(t,\mathbf{x}) &\to - A_0 (-t,\mathbf{x}), \qquad 
   A_i(t,\mathbf{x}) \to  \phantom{+} A_i(-t,\mathbf{x}), \\
     a_0(t,\mathbf{x}) &\to \phantom{+} a_0(-t,\mathbf{x}),
	 \qquad ~ a_i(t,\mathbf{x}) \to - a_i(-t,\mathbf{x}) .
\end{align}
For this symmetry to be consistent with the noncommutative gauge
symmetry, we need to be able to assign transformation laws to the
gauge transformation parameters $\lambda$ and $\alpha$ under PH
conjugation.  While we can do this for $\lambda$ by postulating that
$\lambda(t,\mathbf{x})\to\lambda (-t,\mathbf{x})$, we cannot do the
same for $\alpha$: the two terms in the transformation law for $a_0$
in Eq.~(\ref{a-NCtrans}) always transform differently under PH, no
matter which PH conjugation rule for $\alpha$.
Thus one concludes that emergent noncommutative U(1) gauge symmetry is
fundamentally incompatible with the particle-hole symmetry of the
half-filled Landau level.

There is one more obstacle in writing down a fully noncommutative
description of quantum Hall states.  The general Chern-Simons
description of an abelian quantum Hall state involves, in general,
multiple internal gauge fields with mutual Chern-Simons
terms~\cite{Wen:1992uk}
\begin{equation}
  L \sim \sum_{I,J} K_{IJ} \varepsilon^{\mu\nu\lambda} a^I_\mu \d_\nu a^J_\lambda .
\end{equation}
However, since the NC U(1) symmetry is nonabelian, it is not possible
to write a mutual CS term involving two different NC U(1) gauge
fields.  This problem has been noted recently revisited by Goldman and
Senthil~\cite{Goldman:2021xsm}, where a partial solution, valid to
next order in the expansion over the noncommutativity parameter
$\theta$ is presented.  Here we present a different approach, capable
of solving the problem of the mutual Chern-Simons terms and the
problem of particle-hole symmetry in one scoop, to all orders in the
noncommutativity parameter.

\subsection{Volume preserving diffeomorphism as a ``baby'' version
of noncommutative gauge symmetry}

To solve the problems identified above, we first note a close analogy
between the noncommutative gauge symmetry $U(1)_{\A}$ and volume
preserving diffeomorphism~\cite{Du:2021pbc}\footnote{A nontrivial
connection between area-preserving diffeomorphism and the $W_\infty$
algebra is also found by Dung X.\ Nguyen~\cite{Dung:NC}.}.  We recall
(see Ref.~\cite{Du:2021pbc} for details) that the fractional quantum
Hall problem has invariance under time-dependent diffeomorphisms that
preserve the spatial volume:
\begin{equation}
    x^i \to x^i + \xi^i(t,\mathbf{x}) ,
    \qquad
    \xi^i = \ell^2 \epsilon^{ij} \d_j \lambda ,
\end{equation}
where $\ell^2 = 1/B$ is the magnetic length,
if it is coupled to a scalar potential $A_0$ and a metric $g_{ij}$ that
transform as follows:
\begin{align}
   \delta_\lambda A_0 &= \d_0 \lambda + \theta \{ A_0, \lambda \} , \qquad
   \{ A_0, \lambda \} = \epsilon^{ij} \d_i A_0 \d_j \lambda , \\
   \delta_\lambda g_{ij} &= -\xi^k \d_k g_{ij} - g_{ik} \d_j \xi^k - g_{kj} \d_i \xi^k, \qquad
	\xi^k = \ell^2 \epsilon^{kl} \d_l \lambda .
\end{align}
$A_0$ and $g_{ij}$ are the background fields for volume preserving
diffeomorphisms. Note that the transformation law for $A_0$ is the
long-wavelength limit of that for the noncommutative potential $\A_0$,
Eq.~(\ref{deltaAmu-NC}) Furthermore, one can limit oneself to flat
metrics (which obviously remain flat under diffeomorphisms), which can
be parametrized in terms of the coordinates $\sfX^a(\mathbf{x})$,
defined as the coordinates in which the metric tensor is
$\delta_{ab}$:
\begin{equation}
    g_{ij} = \delta_{ab}\, \d_i \sfX^a \d_j \sfX^b .
\end{equation}
The transformation law for $\sfX^a$ is
\begin{equation}
  \delta \sfX^a = -\xi^i \d_i \sfX^a = \theta \{ \sfX^a, \, \lambda \}.
\end{equation}

If we now write the covariant coordinates as $\mathcal{O}(\theta)$
perturbations around the background coordinates
\begin{equation}
	\sfX^i = x^i +\theta \epsilon^{ij} \sfA_j,
\end{equation}
then the transformation law for the metric induces a transformation
law for the background fields $A_j^b$ which look like
\begin{equation}
   \delta_\lambda \sfA_i = \d_i \lambda + \theta\{ \sfA_i, \, \lambda \} .
\end{equation}
Once again, this is the long-wavelength limit of the transformation
law for the spatial components $\A_i$ of the noncommutative gauge
field.  Combining $A_0$ and $\sfA_i$ into a spacetime vector $\sfA_\mu$
(with $\sfA_0=A_0$), the transformation law of the latter,
\begin{equation}
   \delta_\lambda \sfA_\mu = \d_\mu \lambda
   + \theta \{ \sfA_\mu,\, \lambda \} ,
\end{equation}
is the long wavelength limit of the noncommutative gauge
transformation law where the Moyal brackets are replaced by the
Poisson brackets.  We will call this type of gauge symmetry the ``baby
noncommutative'' (bNC) gauge symmetry.

There exists a well-defined procedure to write down field theories
that are VPD-invariant.  In these theories, each field transforms in a
certain representation (scalar, vector, tensor, etc.) of the VPD
group.  For example, in the Dirac composite fermion theory, the composite
fermion field transforms as
\begin{equation}
  \delta\psi = \theta\{\psi, \, \lambda\}
\end{equation}
(here we ignore the possible coupling of $\psi$ to the spin
connection) and the emergent gauge field transforms as a one-form
under VPD,
\begin{equation}
  \delta_\lambda a_\mu = - \xi^k \d_k a_\mu - a_k \d_\mu \xi^k
  = \theta  \{ a_\mu,\, \lambda \}
    + \theta \epsilon^{kl} a_k \d_\mu\d_l \lambda .
\end{equation}
 We note
that this is \emph{not} what one would expect for the transformation
law of an adjoint field.  Under the internal $U(1)_a$ gauge symmetry,
the transformation law is simply
\begin{equation}
  \delta_\alpha a_\mu = \d_\mu \alpha .
\end{equation}
The absence of the term $\theta\{a_\mu,\, \alpha\}$ allows the
transformation law to be consistent with PH symmetry.

For completeness, we write down the simplest Lagrangian of the
effective field theory of the Dirac composite fermion,
\begin{equation}
  \mathcal L =  \frac i2 v^\mu (\psi^\+ D_\mu \psi - D_\mu\psi^\+ \psi)
  + \frac i2 v_F
   \sigma^a e_a^i (\psi^\+ D_i \psi - D_i\psi^\+ \psi)
   - \frac{a_0}{4\pi\ell^2} + A_0 \left( \frac1{4\pi\ell^2}
   - \frac b{2\pi}\right),
\end{equation}
where (neglecting the coupling of $\psi$ to the spin connection)
$D_\mu\psi = (\d_\mu-ia_\mu)\psi$,
$v^\mu=(1, \ell^2\varepsilon^{ij}\d_j A_0)$, and $e_a^i$ is the
vielbein, which can be defined by $e_a^i\d_i \mathsf{X}^b=\delta_a^b$.
One can check that this theory is invariant under VPDs.

\subsection{SW' map}

We now would like to use the insights obtained from the last Section
to find a way to write down a composite fermion theory that is both
PH-symmetric and can be coupled to an external NC U(1) gauge field
$\A_\mu$.

Our method is inspired by the Seiberg-Witten map between the NC and
commutative gauge symmetries.  We want to derive an analogous map that
maps the NC gauge field $\A_\mu$ to a new gauge field $\bA_\mu$
\begin{equation}\label{bA-A}
  \bA_\mu = \bA_\mu [\A_\mu],
\end{equation}
such that the field $\bA_\mu$ transforms under gauge transformations
as a bNC gauge field, similarly to the field $\sfA_\mu$ discussed in
the previous Section:
\begin{equation}
  \delta \bA_\mu = \d_\mu \blambda + \theta\{ \bA_\mu, \, \blambda \} .
\end{equation}
The gauge parameter $\blambda$ can depend on the gauge field
\begin{equation}
  \blambda = \blambda [\lambda, \A_\mu] .
\end{equation}

If such a map can be found, then the rules for writing down the
coupling of a composite fermion theory to the NC probe $\A_\mu$ is as
follows.  We first couple the theory to the probe $\bA_\mu$ in a way
that respect the bNC gauge symmetry.  Then we restrict ourselves to
flat metrics and parametrize the metric through the coordinates $X^a$,
or equivalently $\bA_i$.  Now we have an action coupled to the baby-NC
gauge fields $\bA_\mu$ and is invariant under baby-NC U(1) gauge
transformations.  We now use the SW' map~(\ref{bA-A}) to replace the
probe field $\bA_\mu$ by the NC gauge field $\A_\mu$.  The result is
an effective theory coupled to the $\A_\mu$ field with NC U(1) gauge
symmetry.

One may wonder if one can use the original Seiberg-Witten map to map
the NC field $\A_\mu$ to a commutative gauge field $\cA_\mu$, which
can be coupled easily to any EFT of the FQHE.  The reason we cannot do
so is the time-dependent shift symmetry~(\ref{A-translation}).  The
map from $\A_\mu$ to $\cA_\mu$ involves $\A_i$ without spatial
derivatives, and a generic coupling of the effective theory to the
commutative field $\cA_\mu$, when rewritten in terms of $\A_\mu$, does
not generally have the shift symmetry~(\ref{A-translation}).

To find the map from the NC theory to the bNC theory, we first develop
a map from the bNC theory to the commutative theory, then use the
known Seiberg-Witten map to map the commutative theory to the NC
theory.  It is possible that there is a more efficient algorithm to
find the map between bNC and NC theories, but we leave finding it to
future work.

\subsection{Map between baby-NC theory and commutative theory}

Suppose we have a commutative theory $\cA_\mu$ which gauge transforms as
\begin{equation}
  \delta \cA_\mu = \d_\mu \clambda.
\end{equation}
We wish to construct a field and a gauge parameter: 
\begin{equation}
  \bA_\mu = \bA_\mu [\cA_\mu], \quad
  \blambda  = \blambda[\clambda, \cA_\mu],
\end{equation}
so that $\bA_\mu$ transforms as a baby-NC gauge field with
parameter $\blambda$,
\begin{equation}\label{bNC-gauge_transf}
  \delta \bA_\mu = \d_\mu \blambda
  +\theta \{ \bA_\mu, \, \blambda \},
\end{equation}
To find these, we will promote $\theta$ to a parameter which we will
call $\tau$ and try to find a family of gauge field $A_\mu(\tau)$
($=A_\mu(\tau)[\cA_\mu]$) and $\lambda(\tau)$
($=\lambda(\tau)[\clambda,\cA_\mu]$) so that for each value of $\tau$
\begin{equation}\label{deltaAtau}
  \delta A_\mu(\tau) = \d_\mu\lambda(\tau)
  + \tau \{ A_\mu(\tau), \, \lambda(\tau) \} .
\end{equation}
Then the commutative and the bNC fields are just $A_\mu(\tau)$ at two
special values of $\tau$: $\cA_\mu=A_\mu(0)$ and
$\bA_\mu=A_\mu(\theta)$.  We differentiate Eq.~(\ref{deltaAtau}) 
over $\tau$ to find that $A_\mu(\tau)$ and $\lambda(\tau)$ must depend
on $\tau$ in such away that the following condition is satisfied:
\begin{equation}\label{theta-evol}
  \delta \dot \bA_\mu(\tau) =
  \d_\mu \dot\blambda(\tau)
  + \{A_\mu(\tau), \, \lambda(\tau) \}
  + \tau \{ \dot A_\mu(\tau), \, \lambda(\tau) \}
  + \tau \{ A_\mu(\tau), \, \dot\lambda(\tau) \}, 
\end{equation}
where dot denotes derivative with respect to $\tau$.  One can check
that this is satisfied if one requires
\begin{subequations}\label{guess}
\begin{align}
  \dot A_\mu(\tau) &=
  \frac12 \vareps^{ij} \big[\d_i A_\mu(\tau) +F_{i\mu}(\tau)  \big] A_j(\tau), \label{SW-bNC}
  \\
  \dot \lambda (\tau) &= \frac12 \vareps^{ij}\d_i \lambda(\tau) A_j(\tau),
\end{align}
\end{subequations}
where $F_{i\mu}(\tau)=\d_i A_\mu(\tau)-\d_\mu A_i(\tau) + \tau
\{A_i(\tau),\, A_\mu(\tau)\}$.
To verify that, one just needs to replace, in Eq.~(\ref{theta-evol}),
$\dot A_\mu(\tau)$ and $\dot\lambda(\tau)$ by the expression given in
Eqs.~(\ref{guess}), then use Eq.~(\ref{deltaAtau}) to convince oneself
that two sides of the equation coincide.

To express the baby-NC field and gauge parameter $\bA_\mu$ and
$\blambda$ in terms of the commutative field $A_\mu$ and gauge
parameter $\lambda$, one solves Eq.~(\ref{SW-bNC}) with the initial
condition $A_\mu(0)=\cA_\mu$, $\lambda(0)=\clambda$.  Then at
$\tau=\theta$ we find the bNC field and gauge parameter:
$A_\mu(\theta)=\bA_\mu$, $\lambda(\theta)=\blambda$.  Inversely, to
express the commutative field $\cA_\mu$ in terms of the bNC field
$\bA_\mu$, one solves this equation with the initial condition at
$\tau=\theta$ and extract the solution at $\tau=0$.

\subsection{Seiberg-Witten map between NC theory and commutative theory}

For convenience we review here the original Seiberg-Witten map between
the NC and commutative field and gauge parameter:
\begin{equation}
  \A_\mu = \A_\mu[\cA_\mu], \qquad \lambda =\lambda[\clambda, \cA_\mu] .
\end{equation}

We now construct a one-parameter family of gauge field and gauge
parameter $\A_\mu(\tau)$ and $\lambda(\tau)$ so that the gauge
transformation law is that of a NC theory with the noncommutativity
parameter $\tau$:
\begin{equation}\label{dncAtau}
  \delta \A_\mu(\tau) = \d_\mu \lambda(\tau)
  + \{\!\!\{ \A_\mu(\tau), \, \lambda(\tau) \}\!\!\}_\tau  ,
\end{equation}
where we introduce the notations
\begin{align}
  (f \star g)_\tau  & = f \exp\left( \frac i2\tau\varepsilon^{ij}\dl_i \dr_j
  \right) g, \\
  \{\!\!\{ f, \, g \}\!\!\}_\tau & =
  \frac1i \left[ (f\star g)_\tau - (g\star f)_\tau \right], \quad
  \{\!\!\{ f, \, g \}\!\!\}^+_\tau = \frac12
  \left[ (f\star g)_\tau + (g\star f)_\tau \right] .
\end{align}
Differentiating Eq.~(\ref{dncAtau}) over $\tau$, we find
\begin{equation}
  \delta \dot \A_\mu(\tau) =
  \d_\mu \dot \lambda(\tau)
  +  \vareps^{ij} \{\!\!\{ \d_i \A_\mu(\tau), \, \d_j \lambda(\tau)\}\!\!\}^+_\tau
  + \{\!\!\{  \dot \A_\mu(\tau), \, \lambda(\tau) \}\!\!\}_\tau
  + \{\!\! \{ \A_\mu(\tau), \, \dot\lambda(\tau) \}\!\!\}_\tau .
\end{equation}
This condition is satisfied if one requires that $\A_\mu(\tau)$ and
$\lambda(\tau)$ satisfy the evolution equations
\begin{subequations}\label{guess3}
\begin{align}
  \dot \A_\mu(\tau) &=
  \frac12 \vareps^{ij} \{\!\!\{ \d_i \A_\mu(\tau) + \F_{i\mu} (\tau), \, \A_j(\tau) \}\!\!\}^+_\tau   ,
  \\
  \dot\lambda(\tau)  &=
  \frac12 \vareps^{ij} \{\!\!\{\d_i \lambda(\tau),\, \A_j(\tau) \}\!\!\}^+_\tau ,
\end{align}
\end{subequations}
where $\F_{i\mu}(\tau)=\d_i\A_\mu(\tau)-\d_\mu\A_i(\tau)+\{\!\!\{ \A_i(\tau), \, \A_\mu(\tau) \}\!\!\}_\tau$.
Solving this system of equations between $\tau=0$ and $\tau=\theta$
one can then establish a mapping between the NC gauge field and a
commutative gauge field.

\subsection{From NC to bNC}

To express the bNC field $\bA_\mu$ and gauge parameter $\blambda$ in
terms of the NC field $\A_\mu$ and gauge parameter $\lambda$, one can
do a two-step process: first one solves Eqs.~(\ref{guess}) for $\tau$
running from $\theta$ to 0 to $\cA_\mu$ and $\clambda$, and then
solves Eqs.~(\ref{guess3}) with $\tau$ running back from 0 to
$\theta$.  The result is an expression relating $\bA_\mu$ and
$\blambda$ with $\A_\mu$ and $\lambda$.
Since Eqs.~(\ref{guess}) and
(\ref{guess}) differ from each other only by terms of order $\theta^2$
and higher on the right-hand sides, the difference between  and
$\bA_\mu$ and $\A_\mu$ starts at order $\theta^3$
\begin{align}
  \bA_\mu &= \A_\mu + \frac{\theta^3}{48}
  \vareps^{ij}\vareps^{i_1j_1}\vareps^{i_2j_2}
  \d_{i_1} \d_{i_2} (2\d_i \A_\mu - \d_\mu\A_i)
  \d_{j_1}\d_{j_2} \A_j + O(\theta^4) ,\\
  \blambda &= \lambda + \frac{\theta^3}{48}
    \vareps^{ij}\vareps^{i_1j_1}\vareps^{i_2j_2}
    \d_{i_1} \d_{i_2} \d_i\lambda\, \d_{j_1}\d_{j_2} \A_j + O(\theta^4) .
\end{align}
Since the difference between $\bA_\mu$ and $\A_\mu$ involves spatial
derivatives of $A_i$, time-dependent shifts on $\A_i$
[Eq.~(\ref{A-translation})] ]simply become time-dependent shifts of
$\bA_i$.

One can now require that fields in the effective field theories of the
FQHE transform as as tensors (or spinor) under VPD with parameter
$\blambda$.  Written in terms of $\lambda$ and $\A$ these
transformation laws are rather complicated and unnatural from the
point of view of noncommutative field theory.  But as the SW' map
demonstrates, there is no requirement that the correct theory of the
FQHE should be a simple noncommutative field theory.

\section{Conclusion}
\label{sec:conclusion}

In this paper we show that the fact that the fact that the quantum
Hall state lives on a single Landau level implies that such a system
can be coupled to an external noncommutative U(1) gauge field in a way
that preserves a NC U(1) gauge symmetry.  We show that the task of
writing down such a theory can be simplified by transforming the NC
U(1) gauge symmetry into a ``baby-NC'' U(1) gauge symmetry, which is
isomorphic to volume-preserving diffeomorphism.  The task of enforcing
VPD invariance in an effective field theory can be accomplished quite
easily with existing tools, for example using the Newton-Cartan
formalism~\cite{Son:2013rqa,Du:2021pbc}.

It appears from the discussion above that the NC U(1) gauge symmetry
does not place any additional constraint on the effective field theory
of the FQHE besides those which can be seen from the VPD.

It would be nice to find a direct map between the NC and bNC theories,
bypassing the need to go through the intermediate commutative theory.
Another remaining open question is on the precise relationship between
the U(1) noncommutative gauge symmetry of the effective field theory
and the GMP algebra~\cite{Girvin:1986zz}.  We leave these questions to
future work.

\acknowledgments

While this paper was being completed, the authors become aware of
Ref.~\cite{Goldman:2021xsm}, where, in particular, the problem with
the mutual Chern-Simons term is discussed and partially solved.  This
paper is supported, in part, by the U.S.\ DOE grant No.\
DE-FG02-13ER41958, a Simons Investigator grant and by the Simons
Collaboration on Ultra-Quantum Matter, which is a grant from the
Simons Foundation (651440, DTS).

\bibliography{noncomm}

\end{document}